\documentstyle[12pt]{article}
\begin{document}
\medskip

\pagestyle{empty}

\rightline{         MRI-PHY/95/10   }

\vspace{5 mm}

\begin{center}

  {\bf SCHRODINGER EQUATION FOR LAGRANGIAN PATH INTEGRAL WITH
  SCALING OF LOCAL TIME}
                     \vspace{1.5 cm}

             $  {\rm   A.K. Kapoor }^\dagger $ \\
             Mehta Research Institute, \\
             10 Kasturba Gandhi Road Allahabad 211002
             INDIA \\
\vspace{3 mm}
                                         and

\vspace{3 mm}

              Pankaj Sharan \\
              Physics Department, Jamia Millia Islamia, \\
              Jamia Nagar, New Delhi 110025, INDIA \\

\vspace{2.5 cm}
                              {\bf ABSTRACT }
\end{center}

     A  method  for  deriving  the Schrodinger equation for Lagrangian
     path integral with scaling of local time is given.

\vspace{2.5 cm}
\hrule
                \noindent
$^\dagger $ Permanent Address : University of Hyderabad, Hyderabad 500134,
INDIA   \\
\ \  email : ashok@mri.ernet.in \\

\newpage
\pagestyle{plain}

\noindent
{\bf 1. Introduction }

\vspace{0.5 cm}

\noindent
     {\it  Scaling of time and exact path integration  } : In 1979 Duru
     and  Kleinert's  important paper [1] on the exact solution for the
     H-atom problem opened the way to the exact path integral treatment
     of  several  potential  problems.  Prior  to  the work of Duru and
     Kleinert the exact  solution to only a few problems could be given
     within   the  path   integral  scheme  [2].  Though  the  original
     treatment   of   the   H-   atom   by   Duru  and  Kleinert  using
     reparametrization  of  paths  was  a formal one, it did lead to an
     important  new  technique  for exact path integration. Besides the
     use  of  technique of adding new  degrees of freedom with  trivial
     dynamics  and  of point transformations, scaling of local time was
     utilized  to  complete exact path integration for several problems
     of quantum mechanics [3-14].

     Let  us  consider  a classical system described by a hamiltonian $
     H(q,p) $. with the Hamilton's equations of motion

\begin{eqnarray}
 \frac{d q^k}{d t} & =& \frac{\partial H}{\partial p}  \\
 \frac{d p^k}{d t} & = & - \frac{\partial H}{\partial q}
\end{eqnarray}

      The  solutions  of these equations give $q$ and $p$ as functions of
      $t$  and  initial  values  $q_0  ,  p_0  $  taken by $q$ and $p$ at
      specified  time $t_0.$ The two functions $q\ \equiv  q(t),$ and $ p
      \equiv   p(t)$  describe  the  classical trajectory in phase space.
      Thus   the  time also plays the role of parameter specifying points
      on the classical paths. The classical path can also be specified in
      terms  of a new parameter $\sigma = \sigma(t)$, such that $\sigma $
      increases  monotonically  with  time; in this paper $ \sigma $ will
      called pseudo-time and could be a different function for each path.
      For  the   present we shall consider only those cases whre $ d t/ d
      \sigma  $  is  a  function of co-ordinates only, say, $\alpha (q)$.
      This  reparametrization  of  paths  can be described by means of an
      auxiliary hamiltonian ${\cal H}$ given by

      \begin{equation}
      {\cal H} = f(q) (H-E)
      \end{equation}

      \noindent
      Imposing  the  energy  constraint $H(q,p) -E =0 $ on the solutions,
      the Hamiltons equations for ${\cal H}$ become
\begin{eqnarray}
     \frac{\partial q^k}{\partial \sigma}
                &  = & \alpha(q) \frac{\partial {\cal H}}{\partial p}  \\
     \frac{\partial p^k}{\partial \sigma}
                & = & - \alpha(q)  \frac {\partial{\cal H}}{\partial q}
\end{eqnarray}
      These  equations  along  with  $ d t/ d\sigma =\alpha(q) $ give the
      correct equations for the classical motion of the particle.

      Duru  and  Kleinert's  treatment  of  hydrogen atom makes  use of
      formal arguments involving reparametrization of paths in the path
      integral  framework of quantum mechanics and Kustarnheimo-Stiefel
      transformation [16]. They succeeded in relating the path integral
      for  the  H-  atom  to  that  for the harmonic oscillator in four
      dimensions.   Subsequent  work  by  others  authors  led  to  the
      important  technique  of scaling of local time. In this technique
      a  central  role  is  played  by  an  identity, which we call the
      scaling formula. For two different problems, this formula relates
      path  integrals  for  the propagators or for the energy dependent
      Green  functions  [17].  The  two path integrals appearing in the
      scaling  formula correspond to the dynamical Hamiltonian $H(q,p)$
      and  an auxiliary hamiltonian ${\cal H}= \alpha (q)( H(q,p) - E +
      \Delta  V  )  $.  Note that in the quantum mechanical problem the
      auxiliary  hamiltonian  differs  from the corresponding classical
      expression  given in (3) above; the difference being by $ O(\hbar
      ^2  )$ terms written as $\Delta V.$ The exact form of $ \Delta V$
      depends on details of the path integral scheme used.

\vspace{0.5 cm}
     \noindent
      {\it Scaling and Hamiltonian Path Integral Quantization  } :
      In  the  exact  treatment  of path integrals the scaling of local
      time is merely used as a tool to arrive at the solution. The idea
      of  scaling  of  local  time  in  path integration has been found
      useful in another way.

      When one wants to set up a hamiltonian path integral quantization
      scheme  in  arbtirary  co-ordinates,  the  path integral with the
      classical  hamiltonian  does  not  lead  to  correct quantization
      rules.  To  recover  the  correct Schrodinger equation one is, in
      general,   forced   to  add,  in  an  ad  hoc  manner,  terms  of
      $O(\hbar^2)$  to  the  classical  hamiltonian.  The  necessity of
      adding  ad-hoc  $O(\hbar^2)  $  terms  is an unwelcome feature of
      hamiltonian  path  integral  quantization  schemes. Even here the
      idea  of  scaling  of  local  time turns out to be useful. It was
      first shown in ref. [18] that it is possible to avoid the need of
      addition  of  these  ad-hoc  $O(\hbar^2)$ terms, and to formulate
      hamiltonian  path  integral  quantization  entirely  in  terms of
      classical  hamiltonian provided one uses a suitable local scaling
      of time [18]. This is a distinct advantage over other hamiltonian
      path   integral  schemes  and  this  once  again  underlines  the
      importance  of  scaling  of  local  time  in   the  path integral
      framework.

\vspace{0.5 cm}
      \noindent
      {\it   Path  Integrals  with  local  scaling  of  time  }  :  The
      hamiltonian  path  integral   quantization  of  ref  [6] has been
      further  developed  in  later  papers  [19-20].  Motivated by the
      importance  of  local  scaling  of  time  in  the  path  integral
      framework,  we  have  introuced  and  studied  properties  of the
      hamiltonian path integrals with local scaling of time. These path
      integrals  were  called  the  path  integrals  of  second kind as
      opposed to the standard path integral which we shall refer to the
      path integral of first kind. The path integral of second type was
      defined  in  terms  of a path integral of first type for the {\it
      auxiliary hamiltonian} as introduced above.

      One  of the most important properties of the path integrals, with
      or  without  a  scaling  of  time  is  the  Schrodinger  equation
      satisfied  by  it.  The  method  for  obtaining  the  Schrodinger
      equation  for the standard path integral  is very simple and well
      known  [9].  For  the  path integral of second kind the method to
      derive   Schrodinger   equation  has  been  given  only  the  for
      hamiltonian  form of path integrals in [19-20]. The derivation of
      the  Schrodinger  equation   for  HPI2  turns  out  to  be rather
      complicated as compared to that for HPI1

      The aim of this paper is to introduce the lagrangian form of path
      integral  with  scaling  and to establish the method of obtaining
      the  Schrodinger equation for this second form of lagrangian path
      integrals.

\newpage

\noindent
     {\bf 2.   Path Integrals with scaling with scaling }

\vspace{0.5 cm}
      In  this  section  we  shall briefly recall the definition of the
      hamiltonian  path  integral  of  second  type  and  to  introduce
      lagrangian path integral with scaling.

      Given  a  hamiltonian  $H(q,p)$  and  a positive scaling function
      $\alpha(q)$,  we  first  define  an  auxiliary hamiltonian ${\cal
      H}(E;q,p) = \alpha(q)(H(q,p) -E) $. The hamiltonian path integral
      of  second kind, denoted by ${\cal K}[H,\alpha,\rho]$, depends on
      the hamiltonian $H(q,p),$ local scaling function $\alpha(q),$ and
      integration measure $\rho$; ${\cal K}$ is defined in terms  of an
      ordinary  hamiltonian  path  integral  $K[{\cal H},\rho]$ for the
      auxiliary hamiltonian ${\cal H}$,  by means of the equation
\begin{eqnarray}
      \lefteqn{  {\cal K}[H,\rho,\alpha] (qt,q_0 0)    } \nonumber
      \\  &&  \equiv \sqrt{\alpha (q)\alpha (q_0)}
      \int \frac{dE}{2 \pi \hbar} \exp(-iEt/\hbar) \int d\sigma
      K[{\cal H}(E;q,p),\rho]
      (q\sigma ,q_00) \hspace{.2 in}    \label{(72)}
\end{eqnarray}

\noindent
     This  path integral ${\cal K}$ was called HPI2, or the hamiltonian
     path  integral of the second kind. It depends on the hamiltonian $
     H(q,p),  $   integration measure $ \rho $  and scaling function  $
     \alpha  (q).  $  In  the  right  hand side, the definition of HPI2
     involves  another  hamiltonian  path  integral HPI1 $ K[\alpha (H-
     E),\rho] $ for the auxiliary hamiltonian

     For  a  phase  space  function  $H(q,p)$  and  integration measure
     $\rho$,    HPI1  has  been  defined in terms of what we have named
     short  time  propagator,  to  be called STP hereafter, and will be
     denoted by $(q_2 \epsilon \Vert q_1 0 ).$  In terms of the STP the
     path  integral  $K[H,\rho]$  is  given  by  means of the following
     equations.
\begin{equation}
      K ^{(N)}[H,\rho](qt;q_0 t_0 ) = \int \prod^{N-1} _{k=1}
        \rho (q_k )dq_k
        \prod ^{N-1}_{j=0} (q_{J+1} \epsilon \Vert q_j 0)      \label{(9.23)}
\end{equation}
where
\begin{equation}
      K [H,\rho ] (qt;q_0 t_0 ) =   \lim _{N \rightarrow \infty }
                   K^{(N)}[H,\rho ](qt;q_0 t_0 )       \label{(9.24)}
\end{equation}

     We  make  a few remarks about the definition of HPI1 and HPI2. The
     STP  approximates  the  full  path  integral $K[H,\rho]$ for short
     times.  The  path  integral  HPI1  is  very  similar  to any other
     hamltonian  path  integral existing in literature  within the time
     slicing  approach  to  the  path  integration.  Finally, the above
     definition  of HPI2  is such that for $\alpha(q) = $ constant, the
     HPI2   $   {\cal    K}[H,\rho,\alpha]    $   coincides  with  HPI1
     $K[H,\rho].$  Further  details of the  definition of HPI1 and HPI2
     and about the choice of STP can be found in our earlier papers[19-
     20].

     We  shall  now  define  lagrangian path integral with scaling in a
     fashion  similar  to the definition of HPI2. We shall restrict our
     attention  to  simple  potential  problems  in  one dimension; the
     generalization  of  our  method  to  many  degrees  of  freedom is
     straightforward.  For  systems  with one degree of freedom we take
     the dynamical hamiltonian to be
\begin{equation}
     H(q,p) =  \frac{p^2}{2M} + V(q)
\end{equation}
     and the auxiliary hamiltonian is
\begin{equation}
          {\cal H } (q,p) = \alpha(q) \left( \frac{p^2}{2M} + V(q) -E )
                                    \right)
\end{equation}
     The  corresponding  dynamical and auxiliary  lagrangians are given
     by
\begin{eqnarray}
     L &=& \frac{M}{2}\left(\frac{\partial q}{ \partial t }\right)^2  -
     V(q)  \\
     {\cal  L}  &  =  &  \frac{M}{2 \alpha(q)}\left( \frac{\partial q}{
     \partial t } \right)^2  - \alpha(q)( V(q) -E )
     \end{eqnarray}

     In  an approach similar to the hamiltonian path integrals we shall
     at  first  define a discrete lagrangian form for the path integral
     with  scaling.  This path integral will be deoted by $ {\cal K}[L,
     \alpha, \rho] $ and is defined by
\begin{eqnarray}
      \lefteqn{  {\cal K}[L,\rho,\alpha] (qt,q_0 0)    } \nonumber
      \\  &&  \equiv \sqrt{\alpha (q)\alpha (q_0)}
      \int \frac{dE}{2 \pi \hbar} \exp(-iEt/\hbar) \int d\sigma
      K[{\cal L},\rho] (q\sigma ,q_00) \hspace{1. in}    \label{(72a)}
\end{eqnarray}
     where the quantity $K[{\cal L},\rho] (q\sigma ,q_00)$ appearing in
     the above equation is an ordinary lagrangian path integral for the
     auxiliary  lagrangian  ${\cal  L}$  defined  below within the time
     slicing approach.

     In  order  to  define  the propagator $ K[{\cal L},\rho]$ we start
     with  the  definition  of  STP  $(q_{k+1},  \epsilon  \vert q_k) $
     corresponding to ${\cal L}$ with measure $\int \rho(q) d^n q,$ for
     short  pseudo  time  $\epsilon = \sigma/N.$. Define
     \vspace{1 cm}
\begin{eqnarray}
     (q_{k+1}, \epsilon \vert q_k)  \hspace{4.3 in} \\
      = \frac{1} { \sqrt {\rho (q_{k+1}) \rho ({q_k}) } }
         \left(\frac{M} {2\pi i\hbar \epsilon f_k} \right)^{1/2}
         \exp \left[ \frac{i}{\hbar} \left(   \frac{\Delta^2_k}{ (2 M
          \epsilon f_k) }
     -    \epsilon f_k(V_k-E )\right)\right]
\label{135}
\end{eqnarray}
\begin{eqnarray}
     f_k & = & f \left( \frac{q_{k+1} + q_k}{2}  \right) \label{136}  \\
     \Delta_k & =&  q_{k+1} - q_k                \label{137}  \\
     q_N & =&  q
\end{eqnarray}
      To  simplify  our  notation,  as in the above equations, we shall
      use  the notation $f_k$ to denote the mid point value $f((q_{k+1}
      +q_k)/2)  $  of  the  function  $  f.  $;  we shall use a similar
      notation for other functions of $q$.   The path integral $K[{\cal
      L},\rho]  (q\sigma  ,q_00)$  is  defined  in  terms  of  the  STP
      $(q_{k+1} \epsilon \vert q_k 0). $

\begin{equation}
      K[{\cal  L},\rho] (q\sigma ,q_00)   = \lim_{N \rightarrow \infty}
      \int \left( \prod^{N-1}_{j=1}
     dq_j\right) \prod^{N-1}_{k=0} \ (q_{k+1} \epsilon \vert q_k 0)
\end{equation}

     In  the above and in the rest of this paper $q_N $ stands for $q$.
     The method for deriving the Schrodinger equation for path integral
     HPI1  is  well  known [21]. In fact evaluation of path integral of
     limit  $  N  \rightarrow  \infty  $ is needed. This is because for
     short  times  HPI1  is  approximated by the STP and this makes the
     derivation  of  the  Schrodinger  equatuion  for HPI1 very simple.
     However,   the  derivation  of  Schrodinger  equation for HPI2  is
     rather  complex  and  was  given  in  our  previous paper. This is
     because in case of HPI2 the STP cannot  be used  in the right hand
     side   to  derive  the  Schrodinger  equation.Expression  of  HPI2
     requires  use  of  full  HPI1 even for short times due to the fact
     that   HPI2 is defined in terms of HPI1 integrated over $\sigma $.
     Therefore  one  has  to insert expressions (14) and (18) into (13)
     and  take  limit  $  N  \rightarrow  \infty  $ at the end. For the
     hamiltonian  path integral with scaling the method of deriving the
     Schrodinger equation is given in our  previous paper.

     \vspace{1 cm}
     {\bf 3. Schrodinger equation }

\vspace{0.5 cm}
     We shall now give the corresponding method for the lagrangian path
     integral  with  scaling.  We use this result to obtain the scaling
     formula  which expresses the lagrangian path integral with scaling
     to lagrangian path integral without scaling. In the following  the
     results are stated and proved  only for $\rho =1 $.

\noindent
      {\it Proposition} : The lagrangian path integral ${\cal K}[ {\cal
      L}, 1, f] $ satisfies
\begin{equation}
     \lim_{t \rightarrow 0}  {\cal K}[ {\cal L}, 1, f](qt,q_00) =
     \delta (q-q_0)
\end{equation}
      and the wave function $\psi(q) \in  L^2(R,dq)$  propagated by the
     path integral
\begin{eqnarray}
     \lefteqn{{\cal K}[ {\cal L}, 1, f](qt,q_00) } \hspace{4 in}
         \nonumber \\
          = \sqrt{f(q)f(q_0)} \int \frac{dE}{2 \pi \hbar} \exp(-iEt/\hbar)
               \int^{\infty}_{0} d\sigma K[{\cal L},\rho] (q\sigma ,q_00)
 \label{139}
\end{eqnarray}
     satisfy the Schrodinger equation
\begin{equation}
     i \hbar \frac{\partial \psi}{\partial t} = - \frac{\hbar^2}{2M}
     \frac{\partial^2 \psi}{\partial q^2}   + (V + v_{ps}) \psi \label{140}
\end{equation}
     where
\begin{equation}
     v_{ps} = -\frac{\hbar^2}{8M} \left\{ (f^"/f) - (f' /f)^2 \right\}
                                                           \label{141}
\end{equation}
     is the one dimensional Pak-Sokmen potential.

\vspace{0.5 cm}
\noindent
     {\it  Outline   of   proof   :}  \/   The   integral  over $E $ is
     performed first. The coefficient  of  $(-iE/\hbar)$   is  $(  t  -
     \sum   f_k)$   which   gives   a   delta   function  $  ë(t-  \sum
     f_k).$   The   $å$ integral  can  be  done  readily, remembering $
     \epsilon=\sigma/N.$  The  net  effect  of  this  is  to  replace $
     \epsilon $ everywhere  by

\begin{equation}
     \epsilon \rightarrow ( \sum f_m) = \eta N / ( \sum f ))        \label{142}
\end{equation}

\noindent
     with   $\eta  =t/N$   and   there   is  an overall Jacobian factor
     coming   from  $  å$  integration.   This  leaves us with the path
     integral  ${\cal  K}$  of (\ref{139}) as $ N  \rightarrow \infty $
     limit of
\begin{eqnarray}
     K_N = \int \left( \prod^{N-1}_{k=1} dq_j \right) G
     \prod^{N-1}_{k=1}
     \left[  \sqrt{ \frac{M}{(2\pi i \hbar \eta)} } \  F_k \exp
     \left\{ \frac{i}{\hbar} \left( \frac{\Delta ^2_k}{2 \eta M}F^{(k)}
     - \frac{\eta V_k}{F^{(k)}}   \right) \right\}
     \right]                                                  \label{143}
\end{eqnarray}
     with
\begin{eqnarray}
     G & = & N \sqrt{f(q)f(q_0)}/ \left( \sum^{N-1}_{m=0} f_m \right)
                                                             \label{144}  \\
     F^{(k)} & =& (\sum^{N-1}_{m=0}  f_m )/ ( N f_k)             \label{145}
\end{eqnarray}

     The   propagator  ${\cal  K}  $  cannot be calculated for all time
     $  t$ for arbitrary  potential  $V.$  However,  our  aim  here  is
     to   obtain  the  Schrodinger  equation. Though for  this  purpose
     it  is  sufficient  to   calculate ${\cal  K}$ only for small $t,$
     nevertheless the $ \lim_{ N  \rightarrow  \infty}$  limit  has  to
     be computed. What we can do is to anticipate  which  terms   would
     be   $O(t),  O(t^2),$  etc., and keep terms  only   up   to  order
     $t.$  This  can  be  done  by  recalling  that  each   power    of
     $\Delta_k$     should     be    counted    as    being   of  order
     $\sqrt{\epsilon}.$ Thus we start from

\begin{equation}
\psi (q,t) = \lim_{N \rightarrow \infty}
              \int K_N (qt,q_0 0)\psi (q_0,0) dq_0              \label{146}
\end{equation}
\noindent
     and our strategy will be to Taylor  expand  everything  so  as  to
     obtain  the   integrand   in   (\ref{146})  as  a function of $ q,
     \Delta_{N-1},..,\Delta_0 $ and to replace

\begin{equation}
     \left(\prod^{N-1}_{k=1} dq_k \right) dq_0
\end{equation}
by
\begin{equation}
     \prod^{N-1}_{k=0} d \Delta_k \label{148}
\end{equation}

     Next,  keeping only appropriate powers of $\Delta,$ we perform the
     $\Delta $ integrations. Lastly, we take the $N \rightarrow \infty$
     limit. Before we start on this long but straight forward  program,
     there  is  one  small  point  which  can be disposed  right  here.
     Recalling  $\Delta_k=O(\sqrt{\eta}),$ we can ignore the  factors $
     1/F^{(k)} $ in  the potential term because for finite $ N, $

\begin{equation}
\frac{1}{N} \sum_m f_m  \approx \frac{1}{N} \sum _k f_k + 0(\Delta_k)
= f_k + O (\Delta_k)                                         \label{149}
\end{equation}
\noindent
     and   we   need   not keep anything higher than order zero because
     $\eta$   is  already  a  factor  in  the potential term. We obtain
     the same  finite $t$ propagator  whether we keep the $0(\Delta_k)$
     terms coming from (\ref{149})  or not. Here after we shall  ignore
     the   potential  term  altogether because it  will  get  added  to
     the   final   Schrodinger  equation in the standard manner. We now
     perform the Taylor expansion
\newpage
\begin{eqnarray}
     \lefteqn{ \psi (q_0) }\hspace{1 cm} & = &  \psi \left(q - \sum^{N-1}_{k=0}
           \Delta_k \right)
\\
     & = & \psi - \left(\sum^{N-1}_{k=0} \Delta_k \right) \psi'
     +\frac{1}{2} \left( \sum^{N-1}_{k=0}  \Delta_k \right)^2  \psi " +     ...
     \label{151}
\end{eqnarray}

     From now on we shall omit explicit dependence on $q.$ Define

\begin{equation}
\delta_k  =(q_{k+1}+q_k)/2 - q_N ,\hspace{1.5 cm}   k=0,1,...,N-1   \label{33}
\end{equation}

     In terms of $ \delta ' s $ we have the following expansions.

\begin{eqnarray}
     f_k &=& f(q_k + \delta_k)=f+ \delta_k f ' +\frac{1}{2}\delta_k^2 (f'/f)^2
                      + ...                                   \label{154} \\
     1/f_k & =&  f^{-1}\left[ 1-\delta_k f'/f -\frac{1}{2} \delta_k^2 f"/f
     + \delta_k^2 \left( f'/f\right)^2+...\right]
                                                              \label{155}\\
     F^{(k)} & =&  1 + (f'/f) \left( \frac{1}{N} \sum^{N-1}_{m=0}  \delta_m
        - \delta_k \right) \\
     &&   + \frac{1}{2}(f"/f) \left( \frac{1}{N}
        \sum^{N-1}_{m=0}  \delta^2_m - \delta^2_k \right) \\
     && + (f'/f)^2\left( \delta^2_k - \frac{\delta_k}{N} \sum^{N-1}
        _{m=0} \delta_m \right)                \label{157}   \\
     \prod_k \sqrt{ F^{(k)} } & =&  1 + \frac{1}{4}(f'/f)^2 \left[
    \sum \delta^2_k -\frac{1}{N} \left( \sum_k \delta_k \right)^2\right] + ...
                                                            \label{158}  \\
     G &=& 1 -\frac{f'}{f} \left( \frac{1}{2} \sum^{N-1}_{k=0}
         \Delta_k + \frac{1}{N} \sum^{N-1}_{m=0}  \delta_m \right)
                                                             \label{159} \\
     && +   \frac{1}{2N}(f'/f)^2 \sum^{N-1}_{k=0} \Delta_k
     \sum^{N-1}_{m=0}  \delta_m \\
     && +  \frac{1}{N^2} (f'/f)^2 \left( \sum \delta_m \right)^2
          -  \frac{1}{8} (f'/f)^2 \left( \sum \Delta_k \right)^2
                                                           \label{160}  \\
     && +\frac{f"}{f} \left[ \frac{1}{4} \left( \sum \Delta_k\right)^2 -
     \frac{1}{2N} \sum \delta^2_m \right] + .....
                                                            \label{161}
\end{eqnarray}

\noindent
     Substituting these in (\ref{143}), we get the gaussian integral
\begin{equation}
\psi(q,t)  =\lim_{N \rightarrow \infty} \int \prod^{N-1}_{k=0}
            \frac{D\Delta_k}{ \sqrt{2\pi i \hbar \eta }} \exp \left[
            \frac{i}{\hbar} \sum^{N-1}_{k=1} \frac{\Delta^2_k}{2 \eta}
            \right] {\cal Q}(\Delta_k)
                                                            \label{162}  \\
\end{equation}
     where
\begin{equation}
     {\cal Q}(\Delta_k)  =  \left\{ \psi (q)-\psi'(q) \sum \Delta_k
+\frac{1}{2}
               \psi "(q) (\sum \Delta_k)^2 \right\} X_1 X_2 X_3 \label{163}
\end{equation}

\noindent
     where $ X_1,X_2, $ and  $X_3 $ are given by

\begin{eqnarray}
    X_1 &=& 1-(f'/f) \left( \frac{1}{2}\sum \Delta_k + \frac{1}{N}\sum \delta_m
                         \right) \\
   && + (f"/f)\left\{\frac{1}{4} \left(\sum\Delta_k\right)^2 - \frac{1}{2N}
   \sum \delta^2_m \right\}   \\
  && + (f'/f)^2 \left\{ \frac{1}{2N} \left(\sum\delta_k\right)
\left(\sum\delta_m
  \right)
  + \frac{1}{N^2} \left(\sum \delta_m\right)^2 -\frac{1}{8}\left(\sum\Delta_k
  \right)^2 \right\}
                                                             \label{164}  \\
X_2 &=& 1 +\frac{iM}{\hbar}  (f'/f) \sum_k \frac{\Delta^2_k}{2 \eta}
           \left(\sum (\delta_m/N)-\delta_k \right) \\
    &&     + \frac{iM}{2 \hbar} (f"/f)\sum_{k}
           \frac{\Delta^2_k}{2 \eta} \left( \sum \delta^2_m /N - \delta^2_k
\right)
           \\
      &&  + \frac{iM}{\hbar} (f'/f)^2 \sum_k  \frac{\Delta^2_k}{2 \eta}
            \left(\delta^2_k - (\delta_k/N) \sum \delta_m/N \right) \\
       && +  \frac{1}{2} (f'/f)^2 \left\{ \frac{iM}{\hbar} \sum_k
             \frac{\Delta^2_k}{2 \eta} \left(\frac{1}{N} \sum \delta_m -
\delta_k
             \right)
                \right\}^2
                \\
X_3 &=&  1 + \frac{1}{4}(f'/f)^2 \left\{ \sum \delta^2_m -\frac{1}{N}  \left(
                      \sum \delta_m \right)^2\right\} \hspace{4 cm}
\end{eqnarray}

     The   above  integrals  are  fairly  straightforward  even  though
     laborious. The points to remember are as follows.

\begin{enumerate}

     {\item    We need to keep only terms of order $t,$ therefore, a fortiori,
               $O(t)$   terms. }

     {\item    The  $\delta's$  are linear combinations of  $ \Delta's $ given
by
               (\ref{33}) }
     {\item    There  are  $  1/ \eta  $  terms in $ X_2$ which means that $
               O(\Delta^4)$ terms  have to be kept for all such terms. }
\end{enumerate}

\noindent
     With   these   points  in mind, (\ref{143}) can be evaluated.  The
     main  labor  is   in  computing  the summations. For  example,  in
     the  term $ (f"/f)(-\sum  \delta^2_m/N),$

\begin{eqnarray}
     \sum^{N-1}_{m=0} \delta^2_m
          &=& \frac{1}{4} \Delta^2_{N-1} + \sum^{N-2}_{m=0}
          \left(  \sum^{N-1}_{j=m+1}\Delta_j +  \frac{1}{2} \Delta_m \right)^2
                                                                           \\
          & =& \frac{1}{4} \Delta^2_{N-1} + \sum^{N-2}_{m=0}   \left[
          \sum^{N-1}_{j=m+1} \Delta_j +\frac{1}{2} \Delta_m \right] \left[
          \sum^{N-1}_{n=m+1} \Delta_n + \frac{1}{2} \Delta_m \right]
                                                                           \\
          & =&\frac{1}{4}  \Delta^2_{N-1} +\sum^{N-2}_{m=0} \left(
          \frac{1}{4} D^2_m + \Delta_m \sum^{N-1}_{j=m+1} \Delta_j +
          \sum^{N-1}_{j=m+1} \sum^{N-1}_{n=m+1} \Delta_j \Delta_n \right)
                   (167)
\end{eqnarray}
\noindent
     which upon integration over $\Delta's$ will become

 \begin{eqnarray}
     \lefteqn{ \left(\sum_k \Delta^2_k/(2\eta) \right)
          \left(-\sum_m \delta^2_m/N \right) \frac{f"}{f} \psi (q) }
                                                                           \\
     & =& - \frac{i \hbar f"}{Nf} \left[ \frac{1}{4} + \sum^{N-
          2}_{m=0} \left( \frac{1}{4}  +\sum^{N-1}_{j=m+1}  \delta_{jm}+
          \sum^{N-1}_{n=m+1} \sum^{N-1}_{j=m+1}  \delta_{jn}
          \right) \right] \psi (q)
            \\
     & =& - \frac{i \hbar \eta f"}{Nf}  \left( \frac{1}{4} +
          \frac{1}{4} (N-1) + 0 +\sum_{m=0}^{N-2} (N-m-1) \right) \psi (q)
                \\
        & =& - \frac{i \hbar t f"}{fN^2} \left(
                \frac{1}{4} +\frac{1}{4} (N-1)  -
                \frac{1}{2}N(N-1)    + ...\right)\psi (q)
                \\  & = &-  \frac{i \hbar t f" }{2f} \psi  \label{168}
\end{eqnarray}

\noindent
     In   the   last   step  we  have  taken  the limit $ N \rightarrow
     \infty.  $   We   omit    the  remaining  details  and  quote  the
     result,  after reinstating the $ q $ dependence,

\begin{eqnarray}
           \psi (q,t) \approx  \psi (q) +\frac{i \hbar t}{2m}  \psi"(q)
           +\frac{i \hbar t}{8m} \left[ (f'/f)^2-(f"/f) \right] \psi (q)
\end{eqnarray}

\noindent
     which  gives us the Schrodinger equation of the proposition  if
     we recall that  the  potential $ V,$ as we  mentioned  above,
     could  have been kept through all this analysis without any
     change.

\noindent
     {\it  Scaling  Formula  } :  As a byproduct of the above result we
     get the scaling formula
\begin{equation}
     {\cal L}[L, \alpha,1 ] = K[ L - v_{PS} , 1 ]
\end{equation}
     This  result  follows from the fact that the two sides satsify the
     same equation and have the same value at $t=0$.

\noindent
     {\it Concluding remarks } :
     \begin{enumerate}
{\item
     In  this  paper  we  have  defined  Lagrangian  path integral with
     scaling   and   have   established  a  method  for  obtaining  the
     Schrodinger  equation  for  this path integral of second kind only
     for a specific choice of the lagrangian STP, obiviously the method
     is  general  and can be used for any other choice of STP. Also the
     method is applicable to systems with several degrees of freedom.}
{\item
       We  have  checked  that  the  results  obtained  here are are in
     agreement  with  those  obtained for the hamiltonian path integral
     with   scaling   obtained  earlier.This  check  was  performed  by
     establishing  a  correspondance between the lagrangian form of STP
     used  here  and  the  hamiltonian forms of STP used in the earlier
     papers. This correspondance is obtained by  first carrying out the
     momentum  integrations  in  the hamiltonian STP of ref [19-20] and
     making repeated use of the McLaughlin Schulman trick [22].}

{\item
     To  conclude  we  indicate  possible new application of the method
     established  in  this  paper. It is well known that semi-classical
     expression  for  the  propagator,   well  known  as the  van Vleck
     Pauli-DeWitt  formula  [23-24],  gives exact answer for propagator
     in  several  cases[26-28].  Recently  there  has  been  a  renewed
     interest  in  finding  the  class of problems for which  the semi-
     classical  result  or  its  suitable  generalizations  give  exact
     answers [29].

     In  this  connection  it  is  intersting  to  ask  if the class of
     problems  for  which WKB result is exact can be enlarged by use of
     scaling  of  time.  Specifically,  let  as  assume  that the semi-
     classical   propagator   for   the   auxiliary  Lagrangian  ${\cal
     L}(E;\dot{q}^K,  q^k),$  or a suitable generalization, is inserted
     for  $K$  in the right hand side of the definintion of ${\cal K}.$
     One  can  then pose the question what is the class of problems for
     which  a  solution  for  the scaling function $\alpha (q) $ can be
     found  such that  (13) gives an exact answer ?  For investigations
     of  this  type,  the  method  of deriving the Schrodinger equation
     established here will certainly be very useful.   }
     \end{enumerate}

     \noindent
     {\it  Acknowledgment  :} One of us ( A.K.K.) thanks Professor H.S.
     Mani  for  hospitality  extended  to  him  at  the  Mehta Research
     Institute.

\newpage

          \noindent
   {\bf REFERENCES }
   \vspace{1 cm}
\begin{enumerate}
{\item I.H.  Duru  and  H.  Kleinert,  Phys.  Lett.{\bf 84B}, 185 (1979);
       Fortschr. der Phys. {\bf 30}, 401 (1982). }

{\item For early evaluations of non-gaussian path Integrals see,
       D.C. Khandekar, and S.V. Lawande, J. Phys. {\bf A5} 812 (1972);
       {\bf A5}, L57 1972; A. Maheswari, J. Phys {\bf A8}, 1019 (1975);
       D. Peak and A. Inomata, J. Math. Phys. {\bf 10}, 1422(1969). }

{\item  R. Ho and A. Inomata, Phys. Rev. Lett. {\bf 48}, 231 (1982). }
{\item  A. Inomata, Phys. Lett. {\bf 87A}, 387 (1982).}
{\item  A. Inomata, Phys. Lett.{\bf 101A} , 253 (1984);
        F.  Steiner,  Phys. Lett. {\bf 106A}, 363 (1984). }
{\item  H. Kleinert, Phys. Lett. {\bf 116A}, 201 (1986). }
{\item  I. Sokmen, Phys. Lett. {\bf 132A}, 65 (1988).}
{\item  H. Durr and A. Inomata, J. Math. Phys. {\bf 26}, 2231 (1985). }
{\item  P. Y. Cai, A. Inomata, and R. Wilson, Phys. Lett.{\bf 99A}, 117 (1983)
     }
{\item  A. Inomata and M. Kayed, J. Phys. {\bf A18}, L235 (1985).}
{\item  A. Inomata and M. Kayed, Phys. Lett. {\bf 108A}, 117 (1983).}
{\item  J.M. Cai, P.Y. Cai, and A. Inomata, Phys. Rev. {\bf A34}, 4621 (1986) }
{\item  S.V. Lawande and K.V. Bhagat, Phys. Lett. {\bf 131A}, 8 (1988);
        D. Bauch, Nuovo Cim. {\bf 85B}, 118 (1985). }
{\item  N.K.Pak  and I. Sokmen, Phys. Lett {\bf 103A}, 298 (1984);
         Phys. Rev. A30  1629 (1984);
         I.H. Duru, Phys. Lett. {\bf 112A}, 421 (1985);
         I. Sokmen, Phys. Lett. {\bf 115A}, 249 (1986);
         S. Erkoc and R. Sever, Phys. Rev. {\bf D30}, 2117 (1984);
         I. Sokmen, Phys. Lett. {\bf 115A}, 6 (1986) }
{\item  D.C. Khandekar and S.V. Lawande, Phys. Reports {\bf 137}, 115 (1986).}

{\item P. Kustanheimo and E. Stiefel, J. Reine Angew Math. {\bf 218},
       204 (1965). }
 {\item  N. Pak and I. Sokmen, Phys. Rev.{\bf A30}, 1629 (1984);
         J.M. Cai, P.Y. Cai, and A. Inomata, Phys. Rev. {\bf A34}, 4621 (1986)
         and references therein;
         Alice Young and Cecile Dewitt-Morette, Ann. Phys. {\bf 169}, 140
         (1986).}
 {\item  A.K. Kapoor, Phys. Rev. {\bf D30}, 1750 (1984).}
 {\item  A.K.Kapoor and Pankaj Sharan, Hyderabad University Preprint HUTP-86/5}
 {\item  A.K.Kapoor and Pankaj Sharan, Mehta Research Institute Preprint
         MRI-PHY/94/17 . }
  {\item
          R.P. Feynman, Revs. Mod. Phys.  {\bf 20} 365  (1948);
          R.P. Feynman and  A.R.  Hibbs, {\it Quantum  Mechanics  and  Path
          Integral,} McGraw Hill, New York, 1965. }
 {\item  D.W. McLaughlin and L.S. Schulman J. Math. Phys. {\bf 12}, 2520
          (1971) }
 {\item  W. Pauli, {\it Selected Topics in field Quantization,}
          MIT  Press, Cambridge, Mass, 1973. }
 {\item  C.M. DeWitt, Phys. Rev.{\bf 81}, 848 (1951).}
 {\item  J.H. Van Vleck, Proc. Natl. Acad. Sci. USA {\bf 14}, 178 (1928).}

 {\item L. S. Schulman Phys. Rev. {\bf 176} (1976) 1558 }
 {\item J. S. Dowker Ann. Phys. {\bf 62} (1971) 561 }
 {\item  J.S. Dowker, J. Phys.{\bf A3 }, 451 (1970).}

 {\item  A.J. Niemi, Ann. Phys. {\bf 235} (1994) 318 and reference therein. }
 \end{enumerate}
\end{document}